\documentstyle[12pt]{article}

\font\twlgot =eufm10 scaled \magstep1
\font\egtgot =eufm8
\font\sevgot =eufm7
\font\twlmsb =msbm10 scaled \magstep1
\font\egtmsb =msbm8
\font\sevmsb =msbm7
\newfam\gotfam

\textfont\gotfam\twlgot
\scriptfont\gotfam\egtgot
\scriptscriptfont\gotfam\sevgot

\newfam\msbfam
\textfont\msbfam\twlmsb
\scriptfont\msbfam\egtmsb
\scriptscriptfont\msbfam\sevmsb

\def\pBbb{\relax\ifmmode\expandafter\Bb\else\typeout{You cann't use
Bbb in text mode}\fi}
\def\Bb #1{{\fam\msbfam\relax#1}}
\def\thebibliography#1{\bigskip\section*{\centering
References\\}\bigskip\list
  {\arabic{enumi}.}{\settowidth\labelwidth{#1}\leftmargin\labelwidth
    \advance\leftmargin\labelsep
    \usecounter{enumi}}
    \def\newblock{\hskip .11em plus .33em minus .07em}
    \sloppy\clubpenalty4000\widowpenalty4000
    \sfcode`\.=1000\relax}
\newcommand{\Si}{\Sigma}

\def\op#1{\mathop{\fam0 #1}\limits}

\newcommand{\pr}{{\rm pr\,}}

\newcommand{\Id}{{\rm Id\,}}
\def\Ker{{\rm Ker\,}}
\newcommand{\ben}{\begin{eqnarray}}
\newcommand{\een}{\end{eqnarray}}
\newcommand{\be}{\begin{eqnarray*}}
\newcommand{\ee}{\end{eqnarray*}}
\newcommand{\bea}{\begin{eqalph}}
\newcommand{\eea}{\end{eqalph}}

\newcommand{\cE}{{\cal E}}

\newcommand{\cT}{{\cal T}}

\newcommand{\al}{\alpha}
\newcommand{\bt}{\beta}
\newcommand{\la}{\lambda}
\newcommand{\La}{\Lambda}

\newcommand{\F}{\Phi}
\newcommand{\p}{\pi}
\newcommand{\om}{\omega}

\newcommand{\m}{\mu}
\newcommand{\n}{\nu}

\newcommand{\G}{\Gamma}

\newcommand{\th}{\theta}

\newcommand{\si}{\sigma}
\newcommand{\w}{\wedge}
\newcommand{\wt}{\widetilde}
\newcommand{\wh}{\widehat}
\newcommand{\ol}{\overline}
\newcommand{\dr}{\partial}

\newcounter{eqalph}
\newcounter{equationa}
\newenvironment{proposition}[1]{{{\it Proposition #1:}}}{}
\newenvironment{lemma}[1]{{{\it Lemma #1:}}}{}
\newenvironment{definition}[1]{{{\it Definition #1:}}}{}
\newenvironment{proof}{{\bf Proof.}}{}
\newenvironment{remark}{{\bf Remark.}}{}

\newenvironment{eqalph}{\stepcounter{equation}
\setcounter{equationa}{\value{equation}}
\setcounter{equation}{0}

\begin{eqnarray}}{\end{eqnarray}
\setcounter{equation}{\value{equationa}}}

\begin{document}
\hbox{}

\centerline{\bf\large MULTIMOMEMTUM HAMILTONIAN FORMALISM}
\medskip

\centerline{\bf\large IN FIELD THEORY. GEOMETRIC SUPPLEMENTARY}
\bigskip

\centerline{\bf Gennadi Sardanashvily}
\medskip

\centerline{Department of Theoretical Physics, Moscow State University}

\centerline{117234 Moscow, Russia}

\centerline{E-mail: sard@theor.phys.msu.su}
\vskip1cm

The well-known geometric approach to field theory is based on description
of classical fields as sections of fibred manifolds, e.g. bundles with
a structure group in gauge theory. In this approach, Lagrangian and
Hamiltonian formalisms including the multiomentum Hamiltonian formalism
are phrased in terms of jet manifolds
\cite{car,gia,got,kol,kup,6sar,sard,lsar}. Then, configuration and phase
spaces of fields are finite-dimensional. Though the jet manifolds have
been widely used for theory of differential operators, the calculus of
variations and differential geometry, this powerful mathematical methods
remains almost unknown for physicists. This Supplementary to our
previous article \cite{lsar} aims to summarize necessary requisites on
jet manifolds and general connections \cite{man,sard,sau}.
\bigskip

All morphisms throughout are differentiable mappings of
class $C^\infty$. Manifolds are real, Hausdorff,
finite-dimensional, second-countable and connected.

We  use the conventional symbols
$\otimes$, $\vee$ and $\wedge$ for the tensor, symmetric and
exterior products respectively.
 The  interior product (contraction) is  denoted by $\rfloor$.

The symbols $\dr^A_B$  mean the partial derivatives with respect to
coordinates with indices $^B_A$.

Given a manifold $M$ with an atlas of local coordinates $(z^\la)$,
the tangent bundle $TM$ of $M$ (resp. the cotangent bundle
$T^*M$ of  $M$) is provided with the atlas of the induced coordinates
$(z^\la, \dot z^\la)$ (resp. $(z^\la,\dot z_\la)$ relative to the
holonomic bases $\dr_\la$ (resp. $dz^\la$).

 If $f:M\to M'$ is a manifold mapping, by
\[
 Tf: TM\to TM' \qquad
 \dot z'^\la= \frac{\dr f^\la}{\dr z^\al}\dot z^\al,
\]
is meant the morphism tangent to $f$.

By $\pr_1$ and  $\pr_2$, we denote the canonical surjections
\[
 \pr_1:A\times B\to A, \qquad \pr_2:A\times B\to B.
\]

We handle the following types of
manifold mappings $f:M\to M'$ when the tangent morphism $Tf$ to
$f$ meets the maximal rank. These are
immersion, submersion and local diffeomorphism when $f$ is both
immersion and submersion.

Recall that a mapping $f:M\to M'$ is called immersion (resp.
submersion) at a point
$z\in M$ when the tangent morphism $Tf$ to $f$ is injection
(resp. surjection) of
the tangent space $T_zM$ to $M$ at $z$ to the tangent space $T_{f(z)}M'$.
  A manifold mapping $f$ of $M$ is termed immersion (resp. submersion)
 if it is immersion (resp. submersion) at all points of $M$.

A triple $ f:M\to M'$ is called the submanifold (resp. the fibred
manifold) if $f$ is both immersion and injection (resp. both
submersion and surjection) of $M$ to $M'$.
 A submanifold  which also is a
topological subspace is called the imbedded submanifold.
Every open subset $U$ of a manifold $M$ is endowed with the
 manifold structure such that the canonical injection
$i_U: U\hookrightarrow  M$ is imbedding.

\section{Fibred manifolds}

Throughout the work, by $Y$ is meant a fibred manifold
\begin{equation}
\pi :Y\to X \label{1.1}
\end{equation}
over an $n$-dimensional base $X$.
We use symbols $y$ and $x$ for points of $Y$ and $X$ respectively.

The total space $Y$ of a fibred manifold $Y\to X$, by definition,
is provided with an atlas of fibred coordinates
\ben
&& (x^\la, y^i),\qquad
 x^\la\circ\p =x^\la, \nonumber\\
 && x^\la \to {x'}^\la(x^\m), \qquad y^i \to
{y'}^i(x^\m,y^j),\label{1.2}
\een
where $(x^\la$ are coordinates of $X$. They are
compatible with the fibration (\ref{1.1}).

 A fibred manifold $Y\to X$ is called
the locally trivial fibred manifold if there exists
a fibred coordinate atlas of $Y$ over  an open covering
$\{\pi^{-1}(U_\xi)\}$ of $Y$ where $\{U_\xi\}$ is an
open covering of the base $X$. In other words, all points of a fibre
of $Y$ belong to the same coordinate chart.
By a differentiable fibre bundle (or simply a bundle), we
mean the locally trivial fibred manifold (\ref{1.1}) provided with a
family of equivalent bundle atlases
\[
\Psi = \{U_\xi, \psi_\xi \}, \qquad
\psi_\xi :\pi^{-1}(U_\xi)\to U_\xi\times V,
\]
where $V$ is a standard fibre of $Y$. Recall that two bundle atlases are
called equivalent if their union also is a bundle atlas.

If $Y\to X$ is a bundle, the fibred coordinates
(\ref{1.2}) of $Y$ are assumed to
be bundle coordinates associated with a bundle atlas  $\Psi$
of $Y$, that is,
\begin{equation}
y^i(y)=(\phi^i\circ\pr_2\circ \psi_\xi)(y), \qquad \pi (y)\in U_\xi,
\label{1.4}
\end{equation}
 where $\phi^i$ are coordinates of the standard fibre $V$ of $Y$.

Given  fibred manifolds
$Y\to X$ and  $ Y'\to X'$, by a fibred morphism is meant a
fibre-to-fibre manifold mapping $\Phi : Y\to Y'$ over a manifold
mapping $f: X\to X'$.
If $f=\Id_X,$  fibred morphism is termed the fibred
morphism $\op\to_X$  over $X$.

In particular, let $X_X$ denotes the  fibred manifold
$\Id_X: X\hookrightarrow X$.
Given a fibred manifold $Y\to X$, a fibred morphism
$X_X\to Y$ over $X$ is a global section  of $Y\to X$. It is a closed
imbedded submanifold. Let $N$ be
an imbedded submanifold of $X$. A fibred morphism $N_N\to Y$ over
$N\hookrightarrow X$ is called a  section of $Y\to X$ over $N$. For each
point $x\in X$, a fibred manifold, by definition, has a section over
an open neighborhood of $x$.

\begin{remark} In accordance with the well-known theorem, if a fibred
manifold $Y\to X$ has a global
section, every section of $Y$ over a closed imbedded submanifold $N$ of
$X$ is extended to a global section of $Y$
due to the properties which are required of a manifold.
\end{remark}

If fibred morphism $Y\to Y'$ over $X$ is a submanifold,
$Y\to X$ is called the fibred submanifold of $Y'\to X$.
Fibred imbedding and fibred diffeomorphism are usually callled the
monomorphism and the isomorphism respectively.

Given a fibred manifold $Y\to X$, every manifold mapping
$f : X' \to X $ yields the pullback $f^*Y\to X'$ comprising the pairs
\[
\{(y,x')\in Y\times X' \mid \quad \pi(y) =f(x')\}
\]
together with the surjection $ (y,x')\to x'$. Every section
$s$ of the fibred manifold $Y\to X$ defines the corresponding pullback
section
\[
(f^*s )(x') = ((s\circ f)(x'),x'), \qquad x'\in X',
\]
of the fibred manifold $f^*Y\to X'$.

In particular, if  a mapping $f$ is a submanifold, the pullback
$f^*Y$ is called the restriction $Y\mid_{f(X')}$ of the
fibred manifold $Y$ to the submanifold $f(X')\subset X$.

The product of fibred manifolds
$\pi:Y\to X$ and $\pi':Y'\to X$ over $X$, by definition, is the total
space of pullbacks
\[
\pi^*Y'={\pi'}^*Y=Y\op\times_X Y'.
\]

 A composite fibred manifold (or simply a composite
manifold) is defined to be composition of surjective submersions
\begin{equation}
 \pi_{\Si X}\circ\pi_{Y\Si}:Y\to \Si\to X. \label{1.34}
\end{equation}
 It is the fibred manifold $Y\to
X$ provided with the particular class of coordinate atlases:
\begin{equation}
( x^\la ,\si^m,y^i) \label{1.36}
\end{equation}
\[
{x'}^\lambda=f^\lambda(x^\mu), \qquad {\sigma'}^m=f^m(x^\mu,\sigma^n),
\qquad  {y'}^i=f^i(x^\mu,\sigma^n,y^j),
\]
where $(x^\m,\si^m)$ are fibred  coordinates  of the fibred manifold
$\Si\to X$.

In particular, let $TY\to Y$ be the tangent bundle
of a fibred manifold $Y\to X$. We have the composite manifold
\begin{equation}
TY\op\to^{T\pi} TX\to X. \label{1.6}
\end{equation}
Given the fibred coordinates (\ref{1.2}) of $Y$, the
corresponding induced  coordinates
of $TY$ are  $(x^\la,y^i,\dot x^\la, \dot y^i)$.

The tangent bundle $TY\to Y$ of a fibred manifold $Y$ has the subbundle
\[
VY = \Ker T\pi
\]
which is called the vertical tangent bundle of $Y$.
This subbundle is provided with the induced coordinates
$(x^\la,y^i,\dot y^i).$

The vertical cotangent bundle $V^*Y\to Y$ of $Y$, by definition, is the
vector bundle dual to the vertical tangent bundle $VY\to Y$,
it is not a subbundle of $T^*Y$.

With $VY$ and $V^*Y$, we have the following exact sequences of bundles
over a fibred manifold
 $Y\to X$:
 \bea
&& 0\to VY\hookrightarrow TY\op\to_Y Y\op\times_X TX\to 0,
\label{1.8a} \\
&& 0\to Y\op\times_X T^*X\hookrightarrow T^*Y\to V^*Y\to 0.
\label{1.8b}
\eea
For the sake of simplicity, we shall denote the products
\[
Y\op\times_X TX, \qquad Y\op\times_X T^*X
\]
by the symbols  $TX$ and $T^*X$ respectively. Different splittings
\[
Y\op\times_X TX\op\to_Y TY, \qquad V^*Y\to T^*Y
\]
of the exact sequences (\ref{1.8a}) and (\ref{1.8b}), by
definition, correspond to different connections on a fibred manifold $Y$.

At the same time, there is the canonical bundle monomorphism
\begin{equation}
\op\w^nT^*X\op\otimes_YV^*Y\op\hookrightarrow_Y\op\w^{n+1}T^*Y.\label{86}
\end{equation}

 Let $\Phi:Y\to Y'$ be a fibred morphism  over $f$.
The tangent morphism $T\Phi$ to $\F$ reads
\begin{equation}
(\dot{x'}^\la,\dot{y'}^i)\circ T\F =(\dr_\m f^\la\dot
x^\m,\dr_\m\Phi^i\dot x^\m +\dr_j\Phi ^i\dot y^j). \label{1.7}
\end{equation}
It is both a linear bundle morphism over $\Phi$ and a fibred morphism
over the tangent morphism $Tf$ to $f$.
Its restriction to the vertical tangent subbundle $VY$
yields  the vertical tangent morphism
\ben
&&V\Phi:VY\to VY', \nonumber\\
&&\dot{y'}^i\circ V\Phi =\dr_j\Phi^i\dot y^j.\label{2}
\een

Vertical tangent bundles of fibred manifolds utilized in field
theory meet almost always the following simple structure.

 One says that a fibred manifold $Y\to X$ has vertical splitting if
there exists the linear isomorphism
\begin{equation}
\al : VY\op\to_Y Y\op\times_X \overline Y \label{1.9}
\end{equation}
where $\overline Y\to X$ is a vector bundle.
The fibred coordinates (\ref{1.2}) of $Y$ are called
adapted to the vertical splitting (\ref{1.9}) if the
induced coordinates of the vertical tangent bundle $VY$ take the form
\[
(x^\m,y^i,\dot y^i = \overline y^i\circ\al)
\]
 where $(x^\m,y^i,\overline y^i)$ are  bundle coordinates
of $\overline Y$. In this case, transition functions $\dot y^i\to\dot y'^i$
between induced coordinate charts are independent on the coordinates $y^i$.

In particular, a vector bundle $Y\to X$ has the canonical
vertical splitting
\begin{equation}
VY=Y\op\times_X Y. \label{1.10}
\end{equation}
An affine bundle $Y$ modelled on a vector
bundle $\overline Y$ has the canonical vertical  splitting
\begin{equation}
VY=Y\op\times_X\overline Y.\label{48}
\end{equation}
Moreover, linear
bundle coordinates of a vector bundle and affine bundle coordinates of an
affine bundle are always adapted to these canonical vertical splittings.

We shall refer to the following fact.

\begin{lemma}{1.1}
Let $Y$ and $Y'$ be fibred manifolds over $X$
and $\Phi:Y\to Y'$ a fibred morphism over $X$. Let $V\Phi$ be the vertical
tangent morphism to $\Phi$. If $Y'$ admits vertical splitting
$VY'=Y'\times\overline Y,$ then there exists the
linear bundle morphism
\begin{equation}
\ol V\Phi:VY\op\to_Y Y\times \ol Y \label{64}
\end{equation}
over $Y$ given by the coordinate expression
\[
{\ol y'}^i\circ\ol V\Phi =\dr_j\Phi^i\dot y^j.
\]
\end{lemma}

By differential forms (or symply forms) on a fibred manifold, we shall
mean exterior, tangent-valued and pullback-valued forms.

Recall that a tangent-valued $r$-form  on a manifold $M$ is
defined to be a section
\[
\phi = \phi_{\la_1\dots\la_r}^\m
dz^{\la_1}\wedge\dots\wedge dz^{\la_r}\otimes\dr_\m
\]
of the bundle $\op\w^r T^*M\otimes TM.$
In particular, tangent-valued 0-forms are vector fields on $M$.

\begin{remark} There is the 1:1 correspondence between
the tangent-valued 1-forms on $M$ and the linear bundle
morphisms $TM\to TM$ or $T^*M\to T^*M$ over $M$
\ben
&&\th:M\to T^*M\otimes TM,\label{29}\\
&&\th: T_zM\ni t\mapsto t\rfloor\th(z)\in T_zM,\nonumber\\
&&\th: T^*_zM\ni t^*\mapsto \th(z)\rfloor t^*\in T^*_zM.\nonumber
\een
For instance, $\Id_{TM}$ corresponds to the canonical tangent-valued 1-form
on  $M$:
\[
\th_M = dz^\la\otimes \dr_\la, \qquad \dr_\la \rfloor \th_M = \dr_\la.
\]
\end{remark}

Let $\op\La^r{\cT}^*(M)$ be the sheaf of exterior $r$-forms on $M$ and
$\cT(M)$ the sheaf of vector fields on $M$.
Tangent-valued $r$-forms on a manifold $M$ constitute the sheaf
$\op\La^r{\cT}^*(M)\otimes\cT(M)$. It is brought into the
 sheaf of graded Lie algebras with respect to
the Fr\"olicher-Nijenhuis (F-N) bracket.

The F-N bracket is defined to be the  sheaf morphism
\[
 \op\La^r {\cT}^*(M)\otimes\cT(M)\times  \op\La^s
{\cT}^*(M)\otimes\cT(M) \to \op\La^{r+s} {\cT}^*(M)\otimes\cT(M),
\]
\be
&& [\phi ,\si] = -(-1)^{\mid\phi\mid\mid\si\mid}[\si,\phi]
=[\al\otimes u,\bt\otimes v] \\
&&\qquad= \al\wedge\bt\otimes [u,v] +
 \al\wedge {\bf L}_u\bt\otimes v - (-1)^{rs}\bt\wedge {\bf L}_v\al
\otimes u \\
&&\qquad +(-1)^r(v\rfloor \al)\wedge d\bt\otimes u-
(-1)^{rs+s}(u\rfloor\bt)\wedge d\al\otimes v,
\ee
\[
\al\in\op\La^r{\cT}^*(M),\quad \bt\in
\op\La^s{\cT}^*(M), \quad u,v\in \cT(M),
\]
where ${\bf L}_u$ and ${\bf L}_v$ are the Lie derivatives of exterior forms.
We have its coordinate expression
\be
 && [\phi,\si] =
(\phi_{\la_1\dots\la_r}^\nu\dr_\n\si_{\la_{r+1}\dots\la_{r+s}}^\m \\
&& \qquad -(-1)^{rs}\si_{\la_1\dots\la_s}^\nu\dr_\nu\phi_{\la_{s+1}\dots
\la_{r+s}}^\m-r\phi_{\la_1\dots\la_{r-1}\nu}^\m
\dr_{\la_r}\si_{\la_{r+1}\dots\la_{r+s}}^\nu
\\
&& \qquad +(-1)^{rs}s\si_{\la_1\dots\la_{s-1}\nu}^\m\dr_{\la_s}
\phi_{\la_{s+1}\dots\la_{r+s}}^\nu)
dz^{\la_1}\wedge\dots\wedge dz^{\la_{r+s}}\otimes\dr_\m.
\ee
Given a tangent-valued form  $\phi$, the Nijenhuis
differential is defined to be the sheaf morphism
\begin{equation}
d_\phi : \si\mapsto d_\phi\si = [\phi,\si]. \label{33}\\
\end{equation}
In particular, if $\th=u$ is a vector field, we have the
Lie derivative of tangent-valued forms
\be
&& {\bf L}_u\si=[u,\si] =(u^\n\dr_\n\si_{\la_1\dots\la_s}^\m -
\si_{\la_1\dots\la_s}^\n\dr_\n u^\m \\
&& \qquad +s\si^\m_{\la_1\dots\la_{s-1}\nu}\dr_{\la_s}u^\nu)
dx^{\la_1}\wedge\dots\wedge dx^{\la_s}\otimes\dr_\m.
\ee

The Nijehuis differential
(\ref{33}) can be extended to exterior forms $\si$ by the rule
\ben
&&d_\phi\si=\phi\rfloor d\si +(-1)^rd(\phi\rfloor\si)
=(\phi_{\la_1\dots\la_r}^\nu\dr_\nu\si_{\la_{r+1}\dots\la_{r+s}}
\nonumber \\
&&\qquad +(-1)^{rs}s\si_{\la_1\dots\la_{s-1}\nu}\dr_{\la_s}
\phi_{\la_{s+1}\dots\la_{r+s}}^\nu)
dz^{\la_1}\wedge\dots\wedge dz^{\la_{r+s}}. \label{32}
 \een
In particular, if $\phi=\th_M$, the familiar exterior differential
\[
d_{\th_M}\si=d\si
\]
is reproduced.

 On a fibred manifold $Y\to X$, we consider the following
particular subsheafs of exterior and tangent-valued forms:
 \begin{itemize}\begin{enumerate}
\item exterior horizontal forms $\phi : Y\to\op\w^r T^*X$;
\item tangent-valued horizontal forms
\be
 && \phi : Y\to\op\w^r T^*X\op\otimes_Y TY,\\
&& \phi =dx^{\la_1}\wedge\dots\wedge dx^{\la_r}\otimes
(\phi_{\la_1\dots\la_r}^\m\dr_\m +
\phi_{\la_1\dots\la_r}^i \dr_i);
\ee
\item tangent-valued projectable horizontal forms
\[
\phi =dx^{\la_1}\wedge\dots\wedge dx^{\la_r}\otimes
(\phi_{\la_1\dots\la_r}^\m(x)\dr_\m +
\phi_{\la_1\dots\la_r}^i(y) \dr_i);
\]
\item vertical-valued horizontal forms
\be
&&\phi : Y\to\op\w^r T^*X\op\otimes_Y VY,\\
&&\phi =\phi_{\la_1\dots\la_r}^i
dx^{\la_1}\wedge\dots\wedge dx^{\la_r}\otimes\dr_i.
\ee
\end{enumerate}\end{itemize}

Vertical-valued horizontal 1-forms on $Y\to X$ are termed the
soldering forms.

By pullback-valued forms on a fibred manifold $Y\to X$, we mean
the morphisms
 \ben
&&Y\to \op\w^r T^*Y\op\otimes_Y TX, \label{1.11} \\
&&Y\to \op\w^r T^*Y\op\otimes_Y T^*X. \label{87}
\een
Let us emphasize that the forms (\ref{1.11}) are not tangent-valued forms
and the forms (\ref{87}) are not exterior forms on $Y$. In particular,
we shall refer to the pullback $\pi^*\th_X$ of the canonical form
$\th_X$ on the base $X$ by $\pi$ onto $Y$. This is a pullback-valued
horizontal 1-form on $Y\to X$ which we denote by the same symbol
\begin{equation}
\th_X:Y\to T^*X\op\otimes_Y TX, \qquad \th_X
=dx^\la\otimes\dr_\la. \label{12}
\end{equation}

Horizontal $n$-forms on a fibred
manifold $Y\to X$ are called horizontal densities. We denote
\[
\om=dx^1\w\dots\w dx^n.
\]

\section{Jet manifolds}

This Section briefs some notions of the
higher order jet formalism and illustrates them by the ones of first
and second order jet machineries.

We use the multi-index $\La$,
$\mid\La\mid=k$ for symmetrized collections $(\la_1...\la_k)$. By
$\La+\la$ is meant the symmetrized collection $(\la_1...\la_k\la)$.

\begin{definition}{1.2} The $k$-order jet space $J^kY$ of a
fibred manifold $Y\to
X$ is defined to comprise all equivalence classes $j^k_xs$, $x\in X$,
of sections $s$ of $Y$ so that sections $s$ and $s'$ belong to the
same class $j^k_xs$ if and only if
\[
\dr_\La s^i(x)=\dr_\La {s'}^i(x), \qquad 0\leq \mid\La\mid \leq k.
\]
\end{definition}

In other words, sections of $Y\to X$ are identified by the first $k+1$
terms of their Teylor series at points of $X$.

Given fibred coordinates (\ref{1.2}) of $Y$, the $k$-order jet space
$J^kY$  is  provided with atlases
of the adapted coordinates
\[
(x^\la, y^i_\La),\qquad  0\leq \mid\La\mid \leq k.
\]
They bring $J^kY$ into a finite-dimensional smooth manifold
satisfying the conditions which we require of a manifold. It possesses
the composite fibration
\[
 J^kY\to J^{k-1}Y\to ... \to Y\to X.
\]

 In particular, the first order jet manifold (or simply the jet
manifold) $J^1Y$ of $Y$ consists of the equivalence classes
$j^1_xs$, $x\in X$, of sections $s$ of $Y$ so that different sections
$s$ and $s'$ belong to the same  class $j^1_xs$ if and only if
\[
Ts\mid _{T_xX} =Ts'\mid_{T_xX}.
\]
In other words, sections $s\in j^1_xs$  are identified by their values
$s^i(x)={s'}^i(x)$ and values of their partial derivatives
$\dr_\mu s^i(x)=\dr_\mu{s'}^i(x)$ at the point $x$ of $X$.

There are the natural surjections
\ben
&&\pi_1:J^1Y\ni j^1_xs\mapsto x\in X, \label{1.14}\\
&&\pi_{01}:J^1Y\ni j^1_xs\mapsto s(x)\in Y. \label{1.15}
\een
We have the composite manifold
\[
\pi_1=\pi\circ\pi_{01}:J^1Y\to Y \to X.
\]
The surjection (\ref{1.14}) is a fibred manifold. The
 surjection (\ref{1.15}) is a bundle. If $Y\to X$ is a bundle, so is
the surjection (\ref{1.14}).

The first order jet manifold $J^1Y$ of $Y$ is provided with the adapted
coordinate atlases
\ben
&&(x^\la,y^i,y_\la^i),\label{49}\\
&&(x^\la,y^i,y_\la^i)(j^1_xs)=(x^\la,s^i(x),\dr_\la s^i(x)),\nonumber\\
&&{y'}^i_\la = (\frac{\dr{y'}^i}{\dr y^j}y_\m^j +
\frac{\dr{y'}^i}{\dr x^\m})\frac{\dr x^\m}{\dr{x'}^\la}. \label{50}
\een
A glance at the transformation
law (\ref{50}) shows that the bundle $J^1Y\to Y$
is an affine bundle. We call it the jet bundle.
It is  modelled on the vector bundle
\begin{equation}
T^*X \op\otimes_Y VY\to Y.\label{23}
\end{equation}

The second order jet manifold $J^2Y$
 of a fibred manifold $Y\to X$ is defined to be the union of
all equivalence classes  $j_x^2s$ of sections $s$ of $Y\to X$ such that
 sections $s\in j^2_xs$ are identified by their values
and values of their first and second order partial
derivatives at the point $x\in X$. The second order jet
manifold $J^2Y$ is endowed with the adapted coordinates
\[
(x^\la ,y^i, y^i_\la,y^i_{\la\m}=y^i_{\m\la}),
\]
\[
y^i_\la (j_x^2s)=\dr_\la s^i(x),\qquad
y^i_{\la\m}(j_x^2s)=\dr_\m\dr_\la s^i(x).
\]

Let $Y$ and $Y'$ be fibred manifolds over SXS and $\Phi:Y\to Y'$
a fibred morphism over a diffeomorphism $f$ of $X$.  It yields the
$k$-order jet prolongation
\[
J^k\Phi: J^kY\ni j^k_xs\mapsto j^k_{f(x)}(\Phi\circ s\circ f^{-1})
\in  J^kY'
\]
of $\Phi$.
In particular, the first order jet prolongation (or simply the jet
prolongation) of $\Phi$ reads
\[
J^1\Phi : J^1Y	\to J^1Y',
\]
\begin{equation}
J^1\Phi :  j_x^1s\mapsto j_{f(x)}^1(\Phi\circ s\circ f^{-1}),\label{26}
\end{equation}
\[
{y'}^i_\la\circ J^1\Phi=\dr_\la(\Phi^i\circ f^{-1}) +\dr_j(\Phi^iy^j_\la
\circ f^{-1}).
\]
It is both an affine bundle morphism over $\Phi$ and a fibred morphism
over the diffeomorphism $f$.

Every section $s$ of a fibred manifold $Y\to X$ admits the $k$-order jet
prolongation to the section
\[
(J^ks)(x)\op=^{\rm def} j^k_xs
\]
of the fibred manifold $J^kY\to X$. In particular, its
first order jet prolongation to the section $J^1s$ of the fibred jet
manifold $J^1Y\to X$ reads
\begin{equation}
(x^\la,y^i,y_\la^i)\circ J^1s= (x^\la,s^i(x),\dr_\la s^i(x)).\label{27}
\end{equation}

Every  vector field
\[
u = u^\la\dr_\la + u^i\dr_i
\]
 on a fibred manifold $Y\to X$ has the  jet lift  to the projectable
vector field
\ben
&&\overline u =r\circ J^1u: J^1Y\to TJ^1Y,\nonumber \\
&& \overline u =
u^\la\dr_\la + u^i\dr_i + (\dr_\la u^i+y^j_\la\dr_ju^i
 - y_\m^i\dr_\la u^\m)\dr_i^\la, \label{1.21}
\een
on the fibred jet manifold $J^1Y\to X$. To construct it, we use the
canonical fibred morphism
\[
r: J^1TY\to TJ^1Y,
\]
\[
(x^\la,y^i,y^i_\la,\dot x^\la, \dot y^i, \dot y^i_\la)\circ r =
(x^\la,y^i,y^i_\la,\dot x^\la,\dot y^i,(\dot y^i)_\la-y^i_\m\dot x^\m_\la),
\]
where $J^1TY$ is the jet manifold of the fibred manifold $TY\to X$.
In particular, there exists the canonical isomorphism
\begin{equation}
VJ^1Y=J^1VY, \qquad \dot y^i_\la=(\dot y^i)_\la, \label{1.22}
\end{equation}
where $J^1VY$
is the jet manifold of the fibred manifold $VY\to X$ and $VJ^1Y$ is the
vertical tangent bundle of the fibred manifold $J^1Y\to X$.
As a consequence, the jet lift (\ref{1.21}) of a vertical vector field
$u$ on a fibred manifold $Y\to X$ consists with its first order jet
prolongation
\[
\overline u=J^1u=u^i\dr_i +(\dr_\la
u^i+y^j_\la\dr_ju^i)\dr^\la_i
\]
to a vertical vector field on the fibred jet manifold $J^1Y\to X$.

Given a second order jet manifold $J^2Y$ of $Y$, we have (i)
the fibred morphism
\[
r_2: J^2TY\to TJ^2Y,
\]
\[
(\dot y^i_\la, \dot y^i_{\la\al})\circ r_2 =
((\dot y^i)_\la-y^i_\m\dot x^\m_\la, (\dot y^i)_{\la\al} -y^i_\m\dot
x^\m_{\la\al} - y^i_{\la\m}\dot x^\m_\al),
\]
and (ii) the canonical isomorphism
\[
VJ^2Y=J^2VY
\]
where $J^2VY$ is the second order jet manifold of the
fibred manifold $VY\to X$ and $VJ^2Y$
is the vertical tangent bundle of the fibred manifold $J^2Y\to X$.

As a consequence, every  vector field $u$
on a fibred manifold $Y\to X$ admits the second order jet lift  to the
projectable  vector field
\[
\overline u =r_2\circ J^2u: J^2Y\to TJ^2Y.
\]
In particular, if
\[
u = u^\la\dr_\la + u^i\dr_i
\]
is a projectable vector field on $Y$, its second order jet lift reads
\ben
&& \overline u =
u^\la\dr_\la + u^i\dr_i + (\dr_\la u^i+y^j_\la\dr_ju^i
 - y_\m^i\dr_\la u^\m)\dr_i^\la+ \nonumber \\
&& \qquad [(\dr_\al +y^j_\al\dr_j +y^j_{\beta\al}\dr^\beta_j) (\dr_\la
+y^k_\la\dr_k)u^i  -y^i_\m\dot
x^\m_{\la\al} - y^i_{\la\m}\dot x^\m_\al]\dr_i^{\la\al}. \label{80}
\een

Given a $k$-order jet manifold $J^kY$ of $Y$, we have the fibred morphism
\[
r_k: J^kTY\to TJ^kY
\]
and  the canonical isomorphism
\[
VJ^kY=J^kVY
\]
where $J^kVY$ is the $k$-order jet manifold of the
fibred manifold $VY\to X$ and $VJ^kY$ is
the vertical tangent bundle of the fibred manifold $J^kY\to X$.
As a consequence, every  vector field $u$
on a fibred manifold $Y\to X$ has the $k$-order jet lift to the projectable
vector field
\ben
&&\overline u =r_k\circ J^ku: J^kY\to TJ^kY, \nonumber \\
&& \overline u =
u^\la\dr_\la + u^i\dr_i + u_\La^i\dr_i^\La, \nonumber\\
&& u_{\La+\la}^i = \wh\dr_\la u_\La^i - y_{\La+\m}^iu^\m, \label{84}
\een
on $J^kY$ where
\begin{equation}
 \wh\dr_\la  = (\dr_\la + y^i_{\Si+\la}\dr_i^\Si), \qquad
 0\leq \mid\Si\mid \leq k. \label{85}
\end{equation}
The expression (\ref{84}) is the $k$-order
generalization of the expressions (\ref{1.21}) and (\ref{80}).

Algebraic structure of a bundle $Y\to X$ also has the jet
prolongation to the jet bundle $J^1Y\to X$ due to the jet
prolongations of the corresponding morphisms.

If $Y$ is a vector bundle, $J^1Y\to X$
does as well. Let $Y$ be a vector bundle and
$\langle\rangle$ the linear fibred morphism
\be
&&\langle\rangle: Y\op\times_X Y^*\op\to_X X\times{\bf R},\\
&& r\circ\langle\rangle = y^iy_i.
\ee
The jet prolongation of $\langle\rangle$ is the linear
fibred morphism
\be
&&J^1\langle\rangle :J^1Y\op\times_X J^1Y^* \op\to_X T^*X\times{\bf R},\\
&& \dot x_\m \circ J^1\langle\rangle = y^i_\m y_i +y^iy_{i\m}.
\ee
Let $Y\to X$ and $Y'\to X$ be vector bundles and $\otimes$ the bilinear
fibred morphism
\be
&& \otimes :Y\op\times_X Y' \op\to_X Y\op\otimes_X Y', \\
&& y^{ik}\circ\otimes = y^iy^k.
\ee
The jet prolongation of $\otimes$ is the bilinear fibred
morphism
\be
&& J^1\otimes : J^1Y\op\times_X J^1Y'\op\to_X J^1(Y\op\otimes_X Y'), \\
&& y^{ik}_\m\circ J^1\otimes =y^i_\m y^k + y^i y^k_\m.
\ee

If $Y$ is an affine
bundle modelled on a vector bundle $\overline Y$, then $J^1Y\to X$
is an affine bundle modelled on the vector bundle $J^1\overline Y\to X$.

\begin{proposition}{1.3} There exist the following bundle monomorphisms:
\begin{itemize}\begin{enumerate}
\item the contact map
\ben
&&\la:J^1Y\op\to_Y
T^*X \op\otimes_Y TY,\label{18}\\
 &&\la=dx^\la\otimes\wh{\dr}_\la=dx^\la
\otimes (\dr_\la + y^i_\la \dr_i),\nonumber
\een
\item the complementary morphism
\ben
&&\th_1:J^1Y \op\to_Y T^*Y\op\otimes_Y VY,\label{24}\\
&&\th_1=\wh{d}y^i \otimes \dr_i=(dy^i- y^i_\la dx^\la)\otimes\dr_i.
\nonumber
\een
\end{enumerate}\end{itemize}\end{proposition}

These monomorphisms
able us to handle the jets as familiar tangent-valued forms.

The canonical morphisms (\ref{18}) and (\ref{24}) define the bundle
monomorphisms
\ben
&& \wh\la: J^1Y\op\times_X TX\ni\dr_\la\mapsto\wh{\dr}_\la
\in J^1Y\op\times_Y TY, \label{30} \\
&&\wh\th_1: J^1Y\op\times_Y V^*Y\ni
dy^i\mapsto\wh dy^i\in J^1Y\op\times_Y T^*Y,  \label{1.18}
\een
The morphism (\ref{30}) yields the canonical horizontal
splitting of the pullback
\begin{equation}
J^1Y\op\times_Y TY=\wh\la(TX)\op\oplus_{J^1Y} VY,\label{1.20}
 \end{equation}
\[
\dot x^\la\dr_\la
+\dot y^i\dr_i =\dot x^\la(\dr_\la +y^i_\la\dr_i) + (\dot y^i-\dot x^\la
y^i_\la)\dr_i.
\]
Accordingly, the morphism (\ref{1.18})
yields the dual canonical horizontal splitting of the pullback
\begin{equation}
J^1Y\op\times_Y T^*Y=T^*X\op\oplus_{J^1Y} \wh\th_1(V^*Y),\label{34}
\end{equation}
\[
\dot x_\la dx^\la
+\dot y_i dy^i =(\dot x_\la + \dot y_iy^i_\la)dx^\la + \dot y_i(dy^i-
y^i_\la dx^\la).
\]
In other words, over $J^1Y$, we have the canonical horizontal splittings of
the tangent and cotangent bundles $TY$ and $T^*Y$ and the corresponding
splittings of the exact sequences (\ref{1.8a}) and (\ref{1.8b}).

In particular, one gets the canonical horizontal splittings of
\begin{itemize}\begin{enumerate}
\item a projectable vector field
\ben
&& u =u^\la\dr_\la +u^i\dr_i=u_H +u_V \nonumber\\
&& \quad =u^\la (\dr_\la +y^i_\la
\dr_i)+(u^i - u^\la y^i_\la)\dr_i, \label{31}
\een
\item an exterior 1-form
\be
&&\si =\si_\la dx^\la + \si^idy^i\\
&&\quad =(\si_\la + y^i_\la\si_i)dx^\la + \si_i(dy^i-
y^i_\la dx^\la),
\ee
\item a tangent-valued projectable horizontal form
\be
&&\phi = dx^{\la_1}\wedge\dots\wedge dx^{\la_r}\otimes
(\phi_{\la_1\dots\la_r}^\m\dr_\m +
\phi_{\la_1\dots\la_r}^i\dr_i)\\
&&\quad = dx^{\la_1}\wedge\dots\wedge dx^{\la_r}\otimes
[\phi_{\la_1\dots\la_r}^\m (\dr_\m  +y^i_\m \dr_i) \\
&&\quad\qquad+(\phi_{\la_1\dots\la_r}^i - \phi_{\la_1\dots\la_r}^\m
y^i_\m)\dr_i],
\ee
\item the canonical 1-form
\ben
&&\th_Y=dx^\la\otimes\dr_\la + dy^i\otimes\dr_i \nonumber\\
&&\quad =\la + \th_1=dx^\la\otimes\wh{\dr}_\la+\wh d
y^i\otimes\dr_i\nonumber \\
&&\quad\qquad = dx^\la\otimes(\dr_\la + y^i_\la \dr_i)
+ (dy^i-y^i_\la dx^\la)\otimes\dr_i.\label{35}
\een
\end{enumerate}\end{itemize}

As an immediate consequence of the splitting (\ref{35}), we get the
canonical horizontal splitting of the exterior differential
\begin{equation}
d=d_{\th_Y}=d_H+d_V=d_\la + d_{\th_1}. \label{1.19}
\end{equation}
Its components $d_H$ and $d_V$ act on the
pullbacks of horizontal exterior forms
\[
\phi_{\la_1\dots\la_r}(y)dx^{\la_1}\wedge\dots\wedge dx^{\la_r}
\]
on a fibred manifold $Y\to X$ by $\pi_{01}$ onto $J^1Y$.
In this case, $d_H$ makes the sense of the total differential
\be
&& d_H\phi_{\la_1\dots\la_r}(y)dx^{\la_1}\wedge\dots\wedge dx^{\la_r}\\
&&\qquad =(\dr_\m + y^i_\m\dr_i)
\phi_{\la_1\dots\la_r}(y)dx^\m\w dx^{\la_1}\wedge\dots\wedge dx^{\la_r},
\ee
whereas $d_V$ is the vertical differential
\be
&& d_V\phi_{\la_1\dots\la_r}(y)dx^{\la_1}\wedge\dots\wedge dx^{\la_r}\\
&&\qquad =\dr_i
\phi_{\la_1\dots\la_r}(y)(dy^i-y^i_\m dx^\m)\w dx^{\la_1}\wedge\dots\wedge
dx^{\la_r}.
 \ee
If $\phi=\wt\phi\om$ is an exterior horizontal density on $Y\to X$, we have
\[
d\phi=d_V\phi=\dr_i\wt\phi dy^i\w\om.
\]

There exist the following second order generalizations of the
contact map (\ref{18}) and the complementary morphism (\ref{24})
to the second order jet manifold $J^2Y$:
\ben
(i)\quad &&\la:J^2Y\op\to_{J^1Y}
T^*X \op\otimes_{J^1Y} TJ^1Y,\nonumber\\
 &&\la=dx^\la\otimes\wh{\dr}_\la=dx^\la
\otimes (\dr_\la + y^i_\la \dr_i + y^i_{\m\la}\dr_i^\m),\label{54}\\
(ii)\quad &&\th_1:J^2Y \op\to_{J^1Y}T^*J^1Y\op\otimes_{J^1Y} VJ^1Y,\nonumber\\
  &&\th_1=(dy^i- y^i_\la dx^\la)\otimes\dr_i +
(dy^i_\m- y^i_{\m\la} dx^\la)\otimes\dr_i^\m.\label{55}
\een

The contact map (\ref{54}) defines the canonical horizontal
splitting of the exact sequence
\[
0\to VJ^1Y\op\hookrightarrow_{J^1Y} TJ^1Y\op\to_{J^1Y}
J^1Y\op\times_X TX\to 0.
\]

In particular, we get the canonical horizontal splitting of a projectable
vector field $\ol u$ on $J^1Y$ over $J^2Y$:
\ben
&&\ol u=u_H+u_V = u^\la[\dr_\la+ y^i_\la+y^i_{\m\la}] \nonumber \\
&& \qquad +[(u^i-y^i_\la u^\la)\dr_i+(u^i_\m- y^i_{\m\la}u^\la)\dr^\m_i].
\label{79}
\een

Using the morphisms (\ref{54}) and (\ref{55}), we can still obtain the
horizontal splittings  of the canonical tangent-valued 1-form $\th_{J^1Y}$
on $J^1Y$ and, as a result, the horizontal splitting of the exterior
differential which are similar the horizontal splittings (\ref{35}) and
(\ref{1.19}):
\ben
&&\th_{J^1Y}=dx^\la\otimes\dr_\la + dy^i\otimes\dr_i +
dy^i_\m\otimes\dr_i^\m=\la +\th_1, \nonumber \\
&& d= d_{\th_{J^1Y}}=d_\la +d_{\th_1}= d_H +d_V. \label{56}
\een

The contact maps (\ref{18}) and (\ref{54}) are the particular cases of
the monomorphism
\ben
&&\la: J^{k+1}Y\to T^*X\op\otimes_{J^kY} TJ^kY,\nonumber\\
&&\la =dx^\la\otimes(\dr_\la + y^i_{\La+\la}\dr_i^\La), \qquad
0\leq \mid\La\mid \leq k. \label{82}
\een

The $k$-order contact map (\ref{82}) sets up the canonical
horizontal splitting of the exact sequence
\[
0\to VJ^kY\hookrightarrow TJ^kY\to J^kY\op\times_X TX\to 0.
\]
In particular, we get the  canonical horizontal splitting of a
projectable vector field $\ol u$ on $J^kY\to X$ over $J^{k+1}Y$:
\[
\ol u=u_H+u_V= u^\la (\dr_\la + y^i_{\La+\la}\dr_i^\La) + (u^i_\La -
 y^i_{\La+\la})\dr_i^\La, \qquad 0\leq \mid\La\mid \leq k .
\]
This splitting  is the $k$-order generalization of the splittings
(\ref{31}) and (\ref{80}).

\begin{definition}{1.4} The repeated jet manifold
$J^1J^1Y$  is defined to be the first order jet manifold of the fibred jet
manifold $J^1Y\to X$.
\end{definition}

 Given the coordinates (\ref{50}) of $J^1Y$, the repeated jet
manifold $J^1J^1Y$ is provided with the adapted coordinates
\begin{equation}
(x^\la ,y^i,y^i_\la ,y_{(\m)}^i,y^i_{\la\m}).\label{51}
\end{equation}

There are the following two bundles:
\ben
(i) \quad&&\pi_{11}:J^1J^1Y\to J^1Y, \label{S1}\\
&&y_\la^i\circ\p_{11} = y_\la^i,\nonumber\\
(ii) \quad&&J^1\pi_{01}:J^1J^1Y\to J^1Y,\label{S'1}\\
&& y_\la^i\circ J^1\pi_{01} = y_{(\la)}^i.\nonumber
\een

Their affine difference over $Y$ yields the Spencer bundle morphism
\[
\delta=J^1\pi_{01} - \pi_{11} :J^1J^1Y\op\to_Y T^*X\op\otimes_Y VY,
\]
\[
(x^\la ,y^i,\dot x_\la\otimes \dot y^i)\circ\delta =(x^\la,
y^i,y_{(\la)}^i-y^i_\la).
\]
The kernel of this morphism is the sesquiholonomic affine subbundle
\begin{equation}
\wh J^2Y\to J^1Y\label{52}
\end{equation}
 of the bundles (\ref{S1}) and (\ref{S'1}).
This subbundle is characterized by the coordinate condition
$y^i_{(\la)}= y^i_\la.$ It is modelled on the vector bundle
\[
\op\otimes^2 T^*X\op\otimes_{J^1Y} VY.
\]
Given the coordinates (\ref{51}) of $J^1J^1Y$,
the sesquiholonomic jet manifold
$\wh J^2Y$ is provided with the adapted coordinates $(x^\la ,y^i, y^i_\la
,y^i_{\la\m})$.
The second order jet manifold $J^2Y$ is the affine
subbundle of the bundle (\ref{52}) given by the coordinate condition
$y^i_{\la\m}=y^i_{\m\la}.$ It is modelled on the vector bundle
\[
\op\vee^2 T^*X\op\otimes_{J^1Y} VY.
\]

We have the following affine bundle monomorphisms
\[
J^2Y\hookrightarrow \wh J^2Y\hookrightarrow J^1J^1Y
\]
over $J^1Y$ and the canonical splitting
\begin{equation}
\wh J^2Y =J^2Y\op\oplus_{J^1Y} (\op\wedge^2 T^*X \op\otimes_Y VY),
\label{1.23}
\end{equation}
\[
y^i_{\la\m} = \frac12(y^i_{\la\m}+y^i_{\m\la}) +
(\frac12(y^i_{\la\m}-y^i_{\m\la}).
\]

Let $\Phi$ be a fibred morphism of a fibred manifold $Y\to X$ to
a fibred manifold $Y'\to X$ over a
diffeomorphism of $X$. Let $J^1\Phi$ be its first order jet prolongation
(\ref{26}). One can consider the first order jet prolongation $J^1J^1\Phi$
of the fibred morphism $J^1\Phi$.
The restriction of the morphism $J^1J^1\Phi$ to the second order jet
manifold $J^2Y$ of $Y$ consists with
the second order jet prolongation $J^2\Phi$ of a fibred morphism $\Phi$.

In particular, the repeated jet prolongation
$J^1J^1s$ of a section $s$ of $Y\to X$ is a section of the fibred manifold
$J^1J^1Y\to X$. It takes its values into $J^2Y$ and
consists  with the second order jet prolongation $J^2s$ of $s$:
\[
(J^1J^1s)(x)=(J^2s)(x)=j^2_xs.
\]

Given the fibred jet manifold $J^kY\to X$, let us consider the repeated jet
manifold  $J^1J^kY$ provided with the adapted coordinates
$(x^\m, y^i_\La, y^i_{\La\la})$. Just as in the case of $k=2$,
there exist two fibred morphisms of $J^1J^kY$
to $J^1J^{k-1}Y$ over $X$. Their difference over $J^{k-1}Y$ is
the $k$-order Spencer morphism
\[
J^1J^kY\to T^*X\op\otimes_{J^{k-1}Y} VJ^{k-1}Y
\]
where $ VJ^{k-1}Y$ is the vertical tangent bundle of the fibred manifold
$J^{k-1}Y\to X$. Its kernel is the
sesquiholonomic  subbundle $\wh J^{k+1}Y$ of the bundle
$J^{k+1}Y\to Y$. It is endowed with the coordinates $(x^\m,
y^i_\La, y^i_{\La\m})$, $\mid\La\mid=k$.

\begin{proposition}{1.5} There exist the fibred monomorphisms
\begin{equation}
J^kY\hookrightarrow\wh J^kY\hookrightarrow J^1J^{k-1}Y \label{76}
\end{equation}
and the canonical splitting
\begin{equation}
\wh J^{k+1}Y= J^{k+1}Y\op\oplus_{J^kY}(T^*X\w\op\vee^{k-1}T^*X\op\otimes_Y
VY),\label{75}
\end{equation}
\[
(x^\m,y^i_\La, y^i_{\La\m}) =(x^\m,y^i_\La, y^i_{(\La\m)}+y^i_{[\La\m]}).
\]
\end{proposition}

We have the following integrability condition.

\begin{lemma}{1.6} Let $\ol s$ be a section of the fibred manifold
$J^kY\to X$. Then, the
following conditions are equivalent:
\begin{itemize}\begin{enumerate}
\item $\ol s=J^ks$ where $s$ is a section of $Y\to X$,
\item $J^1\ol s:X\to \wh J^{k+1}Y$,
\item $J^1\ol s:X\to J^{k+1}Y$.
\end{enumerate}\end{itemize}
\end{lemma}

Building on Proposition 1.5 and Lemma 1.6,
we now describe reduction of higher order differential operators
to the first order ones.

Let $Y$ and $Y'$ be fibred manifolds over $X$ and $J^kY$ the $k$-order
jet manifold of $Y$.

\begin{definition}{1.7} A fibred morphism
\begin{equation}
\cE: J^kY\op\to_X Y'  \label{70}
\end{equation}
is called the $k$-order differential operator (of class $C^\infty$)
on $Y$. It
sends every section $s$ of the fibred manifold $Y$ to the section
$\cE\circ J^ks$ of the fibred manifold $Y'$.
\end{definition}

\begin{proposition}{1.8} Given a fibred manifold $Y$, every
first order differential operator
\begin{equation}
\cE'': J^1J^{k-1}Y\op\to_X Y'\label{72}
\end{equation}
 on $J^{k-1}Y$ defines the $k$-order differential operator
$\cE=\cE''\mid_{J^kY}$ on $Y$.
Conversely, if a first order differential operator
 on $J^{k-1}Y$ exists, any $k$-rder differential operator (\ref{70}) on
$Y$ can be represented by the restriction $\cE''\mid_{J^kY}$ of some first
order differential operator (\ref{72}) on $J^{k-1}Y$ to the $k$-order jet
manifold $J^kY$.
\end{proposition}

\begin{proof} Because of the monomorphism (\ref{76}) every fibred morphism
$J^kY\to Y'$ can be extended to a fibred morphism
$J^1J^{k-1}Y\to Y'.$
\end{proof}

In particular, every $k$-order differential operator (\ref{70}) yields the
morphism
\begin{equation}
\cE'=\cE\circ\pr_2: \wh J^kY\op\to_X Y' \label{77}
\end{equation}
where $\pr_2:\wh J^kY\to J^kY$ is the surjection corresponding to the
canonical splitting (\ref{75}).
It follows that, for every section $s$ of a fibred manifold $Y\to X$,
we have the equality
\[
\cE'\circ J^1J^{k-1}s= \cE\circ J^ks.
\]
We call $\cE'$ (\ref{77}) the  sesquiholonomic differential operator
and consider extensions of a $k$-order differential operator $\cE$
(\ref{70}) to first order differential operators (\ref{72}) only
through its extension
to the sequiholonomic differential operator (\ref{77}).

Let $\ol s$ be a section of the fibred $(k-1)$-order
jet manifold $J^{k-1}Y\to X$ such
that its first order jet prolongation $J^1\ol s$ takes its values into the
sesquiholonomic jet manifold $\wh J^kY$. In virtue of Lemma 1.6,
there exists a section $s$ of $Y\to X$ such that $\ol s=J^{k-1}s$ and
\begin{equation}
\cE'\circ J^1\ol s= \cE\circ J^ks.\label{S5}
\end{equation}

Let $Y'\to X$ be a composite manifold $Y'\to Y\to X$ where $Y'\to Y$
is a vector bundle. Let the $k$-order differential operator
(\ref{70}) on a fibred manifold $Y$ be a fibred morphism over $Y$.
By $\Ker\cE$ is meant the preimage $\cE^{-1}(\wh 0(Y))$ where $\wh 0$
is the canonical zero section of $Y'\to Y$. We say that a section $s$
of  $Y$ satisfies the corresponding system of $k$-order
differential equations if \begin{equation}
J^ks(X)\subset \Ker\cE. \label{73}
\end{equation}

Let a $k$-order differential operator $\cE$ on $Y\to X$
be extended to a first order differential operator
$\cE''$ on $J^{k-1}Y$. Let $\ol s$ be a section of
the fibred manifold $J^{k-1}Y\to X$. We shall say that $\ol s$ is a
sesquiholonomic solution of the corresponding system of first order
differential equations if
\begin{equation}
J^1\ol s(X)\subset \Ker\cE''\cap \wh J^kY. \label{74}
\end{equation}

\begin{proposition}{1.9} The system of the $k$-order differential equations
(\ref{73}) and the system of the first order differential equations (\ref{74})
are equivalent to each other.
\end{proposition}

\begin{proof} In virtue of the relation (\ref{S5}),
every solution $s$ of the equations (\ref{73}) yields the solution
\begin{equation}
\ol s=J^{k-1}s \label{78}
\end{equation}
of the equations (\ref{74}).
Every solution of the equations (\ref{74}) takes
the form (\ref{78}) where $s$ is a solution of the equations (\ref{73}).
\end{proof}

\section{General Connections}

The most of existent differential geometric methods in field theory is
based on principal bundles and principal connections. We follow the
general notion of connections as sections of jet bundles $J^1Y\to Y$
without appealing to transformation groups.

Given a fibred manifold $Y\to X$,
the canonical horizontal splittings (\ref{1.20}) and (\ref{34}) of the
tangent and cotangent bundles $TY$ and $T^*Y$ of $Y$ over the jet bundle
$J^1Y\to Y$ enable us to get the splittings of the exact sequences
(\ref{1.8a}) and (\ref{1.8b}) by means of a section of this jet bundle.

\begin{definition}{1.10} A first order jet field
(or simply a jet field) on a
fibred manifold $Y\to X$ is defined to be a section $\G$ of the affine jet
bundle $J^1Y\to Y$. A first order connection $\G$ on a fibred manifold
$Y$ is defined to be a global jet field
\ben
&&\G :Y\to J^1Y,\nonumber\\
&&(x^\la ,y^i,y^i_\la)\circ\G =(x^\la,y^i,\G^i_\la (y)).\label{61}
\een
\end{definition}

By means of the contact map $\la$ (\ref{18}), every connection $\G$
(\ref{61})
on a fibred manifold $Y\to X$ can be represented by a projectable
tangent-valued horizontal 1-form $\la\circ\G$
on $Y$ which we denote by the same symbol
\ben
&&\G =dx^\la\otimes(\dr_\la +\G^i_\la (y)\dr_i), \label{37}\\
&&{\G'}^i_\la = (\frac{\dr{y'}^i}{\dr y^j}\G_\m^j +
\frac{\dr{y'}^i}{\dr x^\m})\frac{\dr x^\m}{\dr{x'}^\la}.\nonumber
\een

Substituting a connection $\G$ (\ref{37}) into the canonical horizontal
splittings (\ref{1.20}) and (\ref{34}), we obtain the familiar horizontal
splittings
\ben
&&\dot x^\la\dr_\la +\dot y^i\dr_i = \dot x^\la (\dr_\la
+\G^i_\la\dr_i) + (\dot y^i-\dot x^\la\G^i_\la)\dr_i, \nonumber\\
&&\dot x_\la dx^\la +\dot y_idy^i = (\dot x_\la +\G^i_\la\dot
y_i)dx^\la + \dot y_i(dy^i-\G^i_\la dx^\la)
\label{9}
\een
of the tangent and cotangent bundles $TY$ and $T^*Y$ with
respect to a connection $\G$ on $Y$. Conversely, every horizontal splitting
(\ref{9}) determines a tangent-valued form (\ref{37}) and, consequently, a
global jet field on $Y\to X$.

Since the affine jet bundle $J^1Y\to Y$ is modelled on the vector bundle
(\ref{23}), connections on a fibred manifold $Y$ constitute an affine space
modelled on the linear space of soldering forms on $Y$. It follows that, if
$\G$ is a connection and
\[
\si=\si^i_\la dx^\la\otimes\dr_i
\]
 is a soldering form on a fibred manifold $Y$, then
\[
\G+\si=dx^\la\otimes[\dr_\la+(\G^i_\la +\si^i_\la)\dr_i]
\]
is a connection on $Y$. Conversely, if $\G$ and $\G'$ are
connections on a fibred manifold $Y$, then
\[
\G-\G'=(\G^i_\la -{\G'}^i_\la)dx^\la\otimes\dr_i
\]
is a soldering form on $Y$.

For instance, let $Y\to X$ be a vector bundle. A linear connection
on $Y$ reads
\begin{equation}
\G=dx^\la\otimes[\dr_\la+\G^i{}_{j\la}(x)y^j\dr_i]. \label{8}
\end{equation}

One introduces the following basic forms involving a connection $\G$ and a
soldering form $\si$ on a fibred manifold $Y$:

(i) the curvature of $\G$:
\ben
&&R =\frac12 d_\G\G =\frac12 R^i_{\la\m} dx^\la\wedge dx^\m\otimes\dr_i=
\nonumber \\
&&\quad \frac12 (\dr_\la\G^i_\m -\dr_\m\G^i_\la +\G^j_\la\dr_j\G^i_\m
-\G^j_\m\dr_j\G^i_\la) dx^\la\wedge dx^\m\otimes\dr_i; \label{13}
\een

(ii) the torsion of $\G$ with respect to $\si$:
\ben
&&\Omega = d_\si\G =d_\G\si =\frac 12 \Omega^i_{\la\m} dx^\la \wedge
dx^\m\otimes\dr_i= \nonumber\\
&&\quad (\dr_\la\si^i_\m +\G^j_\la\dr_j\si^i_\m -\dr_j\G^i_\la\si^j_\m)
 dx^\la\wedge dx^\m\otimes \dr_i; \label{14}
\een

(ii) the soldering curvature of $\si$:
\ben
&&\varepsilon=\frac12 d_\si\si=\frac12 \varepsilon^i_{\la\m}dx^\la\wedge
dx^\m\otimes\dr_i= \nonumber\\
&&\quad  \frac12 (\si^j_\la\dr_j \si^i_\m - \si^j_\m\dr_j \si^i_\l)
dx^\la\wedge dx^\m\otimes \dr_i. \label{15}
\een

In particular, the curvature (\ref{13}) of the linear
connection (\ref{8}) reads
 \ben
&& R^i_{\la\m}(y)=R^i{}_{j\la\m}(x)y^j, \nonumber\\
&&R^i{}_{j\la\m}=\dr_\la\G^i{}_{j\m} -\dr_\m\G^i{}_{j\la}
+\G^k{}_{j\la}\G^i{}_{k\m} -\G^k{}_{j\m}\G^i{}_{k\la}.\label{25}
\een

A connection $\G$ on a fibred manifold $Y\to X$ yields the affine
bundle morphism
\ben
&&D_\G:J^1Y\ni z\mapsto z-\G(\pi_{01}(z))\in T^*X\op\otimes_Y VY,
\label{38}
\\
&&D_\G =(y^i_\la -\G^i_\la)dx^\la\otimes\dr_i. \nonumber
\een
It is  called the covariant differential. The corresponding covariant
derivative of  sections $s$ of  $Y$ reads
\begin{equation}
\nabla_\G s=D_\G\circ J^1s=(\dr_\la s^i-
(\G\circ s)^i_\la)dx^\la\otimes\dr_i \label{39}
\end{equation}
A section $s$ of a fibred manifold $Y$ is called an integral section for a
connection $\G$ on $Y$ if
\begin{equation}
\G\circ s=J^1s,\label{40}
\end{equation}
that is, $ \nabla_\G s=0$.

Now, we  consider some particular  properties of linear
connections on vector bundles.

Let $Y\to X$ be a vector bundle and
$\G$ a linear connection on $Y$. On the dual vector bundle $Y^*\to X$,
there exists the dual connection $\G^*$ is called the dual connection
to $\G$ given by the coordinate expression
\[
\G^*_{i\la}=-\G^j{}_{i\la}(x)y_j.
\]

For instance, a linear connection $K$ on the tangent bundle $TX$ of a
manifold $X$ and the dual connection $K^*$ to $K$ on the cotangent bundle
$T^*X$ read
\ben
&& K^\al_\la=K^\al{}_{\nu\la}(x)\dot x^\nu,\nonumber\\
&&K^*_{\al\la}=-K^\nu{}_{\al\la}(x)\dot x_\nu, \label{408}
\een

Let $Y$ and $Y'$ be vector bundles over $X$. Given linear connections
$\G$ and $\G'$ on $Y$ and $Y'$ respectively, the tensor product
connection $\G\otimes\G'$ on the tensor product
\[
Y\op\otimes_X Y'\to X
\]
is defined. It takes the coordinate form
\[
(\G\otimes\G')^{ik}_\la=\G^i{}_{j\la}y^{jk}+{\G'}^k{}_{j\la}y^{ij}.
\]

The construction of the dual connection and the tensor product connection
can be extended to connections on composite manifolds (\ref{1.34}) when
$Y\to\Si$ is a vector bundle.

Let $Y\to\Si\to X$ be the composite  manifold (\ref{1.34}). Let $J^1\Si$,
$J^1Y_\Si$ and $J^1Y$ the first order jet manifolds of
$\Si\to X$, $Y\to \Si$ and $Y\to X$ respectively. Given fibred
coordinates $(x^\la, \si^m, y^i)$ (\ref{1.36}) of $Y$,
the corresponding adapted coordinates of the jet manifolds $J^1\Si$,
$J^1Y_\Si$ and $J^1Y$ are
\ben
 &&( x^\la ,\si^m, \si^m_\la),\nonumber\\
 &&( x^\la ,\si^m, y^i, \wt y^i_\la, y^i_m),\nonumber\\
 &&( x^\la ,\si^m, y^i, \si^m_\la ,y^i_\la) .\label{47}
\een

We say that a connection
\[
A=dx^\lambda\otimes[\dr_\lambda+\Gamma^m_\lambda (\si)
\dr_m + A^i_\lambda (y)\dr_i], \qquad \si=\pi_{Y\Si}(y),
\]
on a composite manifold $Y\to\Si\to X$ is projectable to a connection
\[
\G= dx^\lambda\otimes[\dr_\lambda+\Gamma^m_\lambda (\si)\dr_m]
\]
on the fibred manifold $\Si\to X$, if there exists the commutative diagram
\[
\begin{array}{rcccl}
 & {J^1Y} &  \op\longrightarrow^{J^1\pi_{Y\Si}} & {J^1\Si} &  \\
{_A} &\put(0,-10){\vector(0,1){20}} & & \put(0,-10){\vector(0,1){20}} &
{_\G}
\\
& {Y} & \op\longrightarrow_{\pi_{Y\Si}} & {\Si} &
\end{array}
\]
Let $Y\to \Si$ be a vector bundle and
\[
A=dx^\lambda\otimes[\dr_\lambda+\Gamma^m_\lambda (\si)
\dr_m + A^i{}_{j\lambda}(\si )y^j\dr_i]
\]
a linear morphism over $\G$.
Let $Y^*\to\Si\to X$ be a composite manifold where $Y^*\to \Si$
is the  vector bundle dual to $Y\to \Si$. On $Y^*\to X$, there exists
the dual connection
\begin{equation}
A^*=dx^\lambda\otimes[\dr_\lambda+\Gamma^m_\lambda (\si)
\dr_m - A^j{}_{i\lambda}(\si )y_j\dr^i] \label{65}
\end{equation}
projectable to $\G$.

Let $Y\to\Si\to X$ and $Y'\to\Si\to X$ be composite manifolds
where $Y\to\Si$ and $Y'\to\Si$ are vector bundles. Let
$A$ and $A'$ be connections on $Y$ and $Y'$ respectively
which are projectable to the same connection $\G$ on the fibred
manifold $\Si\to X$.
On the tensor product
\[
Y\op\otimes_\Si Y'\to X,
\]
there exists the tensor product connection
 \begin{equation}
A\otimes A'=dx^\lambda\otimes(\dr_\lambda+\Gamma^m_\lambda
\dr_m+(A^i{}_{j\la}y^{jk}+{A'}^k{}_{j\la}y^{ij})\dr_{ik})\label{66}
\end{equation}
projectable to $\G$.

In particular, let $Y\to X$ be a fibred manifold and
$\G$  a connection on  $Y$. The vertical tangent
morphism $V\G$ to $\G$ defines the connection
\ben
&&  V\G :VY\to VJ^1Y=J^1VY,\nonumber \\
&& V\G =dx^\la\otimes(\dr_\la
+\G^i_\la\frac{\dr}{\dr y^i}+\dr_j\G^i_\la\dot y^j
\frac{\dr}{\dr \dot y^i}), \label{43}
\een
on the composite manifold $VY\to Y\to X$ due to the canonical isomorphism
(\ref{1.22}). The connection $V\G$ is projectable to the connection $\G$ on
$Y$, and it is a linear bundle morphism over $\G$:
\[
\begin{array}{rcccl}
& {VY} &  \op\longrightarrow^{V\G} & {J^1VY} &  \\
&\put(0,10){\vector(0,-1){20}} & & \put(0,10){\vector(0,-1){20}} & \\
& {Y} & \op\longrightarrow^\G & {J^1Y} &
\end{array}
\]
The connection (\ref{43}) yields the  connection
\begin{equation}
V^*\G =dx^\la\otimes(\dr_\la +\G^i_\la\frac{\dr}{\dr
y^i}-\dr_j\G^i_\la \dot y_i \frac{\dr}{\dr \dot y_j}) \label{44}
\end{equation}
on the composite manifold $ V^*Y\to Y\to X$ which is the dual connection to
$V\G$ over $\G$.

Now, we consider second-order connections.

\begin{definition}{1.11} A second order jet field (resp. a second order
connection)  $\ol\G$ on a fibred manifold $Y\to X$ is defined to be
a first order jet field (resp. a first order connection)
on the fibred jet manifold $J^1Y\to X$, i.e.
this is a section (resp. a global section) of the  bundle (\ref{S1}).
\end{definition}

In the coordinates (\ref{51}) of the repeated
jet manifold $J^1J^1Y$, a second
order jet field $\ol\G$ is given by the expression
\[
(y^i_\la,y^i_{(\m)},y^i_{\la\m})\circ\ol\G=
(y^i_\la,\ol\G^i_{(\m)},\ol\G^i_{\la\m}).
\]
Using the contact map (\ref{54}), one can represent it by the horizontal
1-form
\begin{equation}
\ol\G=dx^\m\otimes (\dr_\m
+\ol\G^i_{(\m)}\dr_i+\ol\G^i_{\la\m}\dr^\la_i)\label{58}
\end{equation}
on the fibred jet manifold $J^1Y\to X.$

A second order jet field $\ol\G$ on $Y$ is termed a sesquiholonomic (resp.
holonomic) second order jet field if it takes its values into the subbundle
$\wh J^2Y$ (resp. $J^2Y$) of $J^1J^1Y$. We have the coordinate equality
$\ol\G^i_{(\m)}=y^i_\m$
 for a sesquiholonomic second order jet field and the additional equality
$\ol\G^i_{\la\m}=\ol\G^i_{\m\la}$
 for a holonomic second order jet field.

Given a first order connection $\G$ on a fibred manifold $Y\to X$,
one can construct a second order connection on $Y$, that is,
a connection on the fibred jet  manifold $J^1Y\to X$ as follows.

The first
order jet prolongation $J^1\G$ of the connection $\G$ on $Y$ is
 a section of the bundle (\ref{S'1}), but not the bundle $\pi_{11}$
(\ref{S1}).
Let $K^*$ be a linear symmetric connection (\ref{408}) on the cotangent bundle
$T^*X$ of $X$:
\[
K_{\la\m}=-K^\al{}_{\la\m}\dot x_\al, \qquad K_{\la\m}=K_{\m\la}.
\]
There exists the  affine fibred morphism
\[
 r_K: J^1J^1Y\to J^1J^1Y, \qquad  r_K\circ r_K=\Id_{J^1J^1Y},
\]
\[
(y^i_\la ,y_{(\m)}^i,y^i_{\la\m})\circ r_K=
(y^i_{(\la)} ,y_\m^i,y^i_{\m\la}+ K^\al{}_{\la\m}(y^i_\al
- y^i_{(\al)})).
\]
One can verify the following transformation relations
of the coordinates (\ref{51}):
\be
&&{y'}^i_\m\circ r_k={y'}^i_{(\m)}, \qquad
{y'}^i_{(\m)}\circ r_k={y'}^i_\m, \\
&&{y'}^i_{\la\m}\circ r_k={y'}^i_{\m\la} +{K'}^\al{}_{\la\m}({y'}^i_\al
- {y'}^i_{(\al)}).
\ee
Hence, given a first order connection $\G$ on a fibred
manifold $Y\to X$, we have the second order connection
\[
J\G\op = r_K\circ J^1\G,
\]
\begin{equation}
J\G=dx^\m\otimes [\dr_\mu+\Gamma^i_\mu\dr_i +(\dr_\la\Gamma^i_\m+
\dr_j\Gamma^i_\mu y^j_\lambda -
K^\alpha{}_{\lambda\mu} (y^i_\alpha-\Gamma^i_\alpha))
\dr_i^\la],\label{59}
\end{equation}
on $Y$. This is an affine morphism
\[
\begin{array}{rcccl}
 & {J^1Y} &  \op\longrightarrow^{J\G} & {J^1J^1Y} &  \\
{_{\pi_{01}}} &\put(0,10){\vector(0,-1){20}} &
& \put(0,10){\vector(0,-1){20}} & {_{\pi_{11}}} \\
 & {Y} & \op\longrightarrow_{\G} & {J^1Y} &
\end{array}
\]
over the first order connection $\G$.

Note that the curvature $R$ (\ref{13}) of a first order connection $\G$ on
a fibred manifold $Y\to X$ induces the soldering form
\begin{equation}
\ol\si_R=R^i_{\la\m}dx^\m\otimes\dr^\la_i \label{60}
\end{equation}
on the fibred jet
manifold $J^1Y\to X$.  Also the torsion (\ref{14}) of
a first order connection $\G$ with respect to a soldering
form $\si$ on $Y\to X$ and the soldering curvature (\ref{15}) of $\si$
define soldering forms on $J^1Y\to X$.


\begin{thebibliography}{ederf}

\bibitem{car} J. Cari\~nena, M. Crampin and L. Ibort, {\it Differential
Geometry and its Application} {\bf 1} (1991) 345.

\bibitem{gia} G. Giachetta  and L. Mangiarotti,
{\it International Journal of Theoretical Physics} {\bf 29} (1990), 789.

\bibitem{got} M. Gotay,  in
{\it Mechanics, Analysis and Geometry: 200 Years after Lagrange}, ed.
M.Francaviglia (Elseiver Science Publishers B.V., 1991) p. 203.

\bibitem{gun} C. G\"unther, {\it Journal of
Differential Geometry} {\bf 25} (1987) 23.

\bibitem{kol} I. Kol\'a\v r,  {\it Journal of Geometry
and Physics} {\bf 1} (1984) 127.

\bibitem{kup} B. Kupershmidt, in  Lect.  Notes in Math., Vol.
775 (Springer, Berlin - New York, 1980) p.162.

\bibitem{man} L. Mangiarotti and M. Modugno,  in {\it
Geometry and  Physics}, ed. M. Modugno (Pitagora Editrice, Bologna, 1982)
p. 135.

\bibitem{6sar}
G. Sardanashvily and O. Zakharov, {\it Differential Geometry
and its Applications}  {\bf 3} (1993) 245.

\bibitem{sard} G. Sardanashvily, {\it Gauge Theory in Jet Manifolds}
(Hadronic Press Inc., Palm Harbor, 1993).

\bibitem{lsar} G.Sardanashvily, Multimomentum Hamiltonian Formalism in
Field Theory, LaTeX preprint: hep-th/9403172

\bibitem{sau} D. Saunders, {\it The Geometry of Jet Bundles}
(Cambridge Univ. Press, Cambridge, 1989).


\end{thebibliography}
\end{document}